\documentclass[%
 reprint,
superscriptaddress,
 amsmath,amssymb,
 aps,
prb,citeautoscript
]{revtex4-2}

\usepackage{graphicx}
\usepackage{dcolumn}
\usepackage{bm}
\usepackage{lipsum}
\usepackage{siunitx}
\usepackage{placeins}
\usepackage{hyperref}
\usepackage{lineno}

\usepackage{nameref}
\usepackage{amstext}
\usepackage{mathtools} 
\usepackage{amsmath,amssymb}    
\usepackage{booktabs}
\usepackage{array}
\usepackage{xcolor}
\usepackage{tabularx}
\usepackage{ulem}

\usepackage{booktabs}

\DeclareSIUnit{\rad}{rad}
\DeclareSIUnit{\deg}{deg}

\hypersetup{
        colorlinks   = true,
        citecolor    = blue,
        linkcolor    = blue}

\DeclareSIUnit\bar{bar}
\DeclareSIUnit\torr{Torr}

\makeatletter
\def\@email#1#2{%
 \endgroup
 \patchcmd{\titleblock@produce}
  {\frontmatter@RRAPformat}
  {\frontmatter@RRAPformat{\produce@RRAP{*#1\href{mailto:#2}{#2}}}\frontmatter@RRAPformat}
  {}{}
}%

\let\svthefootnote\thefootnote
\newcommand\freefootnote[1]{%
  \let\thefootnote\relax%
  \footnotetext{#1}%
  \let\thefootnote\svthefootnote%
}

\begin{document}

\preprint{APS/123-QED}

\title{ 
Evolution of dissipative regimes in atomically thin Bi$_2$Sr$_2$CaCu$_2$O$_{8+x}$ superconductor 
}

 \author{Sanaz Shokri}
 \affiliation{Leibniz Institute for Solid State and Materials Science Dresden (IFW Dresden), 01069 Dresden, Germany}
 \affiliation{Institute of Applied Physics, Technische Universität Dresden, 01062 Dresden, Germany}

 \author{Michele Ceccardi}
 \affiliation{University of Genova, Department of Physics, Via Dodecaneso 33, 16146 Genova, Italy}
 \affiliation{CNR-SPIN Institute, Corso Perrone 24, 16152 Genova, Italy}
 
     \author{Tommaso Confalone}
 \affiliation{Leibniz Institute for Solid State and Materials Science Dresden (IFW Dresden), 01069 Dresden, Germany}
  \affiliation{Institute of Applied Physics, Technische Universität Dresden, 01062 Dresden, Germany}

\author{Christian N. Saggau}
 \affiliation{Leibniz Institute for Solid State and Materials Science Dresden (IFW Dresden), 01069 Dresden, Germany}
 \affiliation{DTU Electro, Department of Electrical and Photonics Engineering, Technical University of Denmark, Lyngby, Denmark}
\affiliation{Center for Silicon Photonics for Optical Communications (SPOC), Technical University of Denmark, Lyngby, Denmark}

\author{Yejin Lee}
\affiliation{Max Planck Institute for Chemical Physics of Solids, 01187 Dresden, Germany}
 \affiliation{Leibniz Institute for Solid State and Materials Science Dresden (IFW Dresden), 01069 Dresden, Germany}

 \author{Mickey Martini}
\affiliation{Swabian Instruments GmbH, Stammheimer Str. 41, 70435 Stuttgart, Germany}
 \affiliation{Leibniz Institute for Solid State and Materials Science Dresden (IFW Dresden), 01069 Dresden, Germany}
 
\author{Genda Gu}
 \affiliation{Condensed Matter Physics and Materials Science Department, Brookhaven National Laboratory, Upton, NY 11973, USA}

  \author{Valerii\,M.\,Vinokur}
 \affiliation{Terra Quantum AG, Kornhausstrasse 25, St. Gallen, 9000, Switzerland}

 \author{Ilaria Pallecchi}
\affiliation{CNR-SPIN Institute, Corso Perrone 24, 16152 Genova, Italy}
 
 \author{Kornelius Nielsch}
 \affiliation{Leibniz Institute for Solid State and Materials Science Dresden (IFW Dresden), 01069 Dresden, Germany}
 \affiliation{Institute of Applied Physics, Technische Universität Dresden, 01062 Dresden, Germany}
 \affiliation{Institute of Materials Science, Technische Universität Dresden, 01062 Dresden, Germany}

   \author{Federico Caglieris}
 \affiliation{CNR-SPIN Institute, Corso Perrone 24, 16152 Genova, Italy}
   
 \author{Nicola Poccia}
 \affiliation{Leibniz Institute for Solid State and Materials Science Dresden (IFW Dresden), 01069 Dresden, Germany}
 \affiliation{Department of Physics, University of Naples Federico II, Via Cintia, Naples 80126, Italy}

\begin{abstract}

 Thermoelectric transport has been widely used to study Abrikosov vortex dynamics in unconventional superconductors. However, only a few thermoelectric studies have been conducted near the dimensional crossover that occurs when the vortex-vortex interaction length scale becomes comparable to the sample size. Here we report the effects of finite size on the dissipation mechanisms of the Nernst effect in the optimally doped Bi$_2$Sr$_2$CaCu$_2$O$_{8+x}$ high-temperature superconductor, down to the atomic length limit. To access this regime, we develop a new generation of thermoelectric chips based on silicon nitride microprinted circuit boards. These chips ensure optimized signals while preventing sample deterioration. Our results demonstrate that lateral confinement at the nanoscale can effectively reduce vortex dissipation. Investigating vortex dissipation at the micro- and nano-scale is essential for creating stable, miniaturized superconducting circuits.
\end{abstract}

\maketitle
\section*{Introduction}
The creation of atomically thin  van der Waals (vdW)
superconducting $\text{Bi}_{2}\text{Sr}_{2}\text{CaCu}_{2}\text{O}_{8+x}$ (BSCCO) layers
created new possibilities for the investigation of high temperature topological superconductivity\,\cite{yu2019high, zhao2019sign}, superconducting fluctuations at the nanoscale and the exploration of topological defects\,\cite{hu2024vortex}. The vdW stacking of BSCCO thin crystals offers a unique platform for engineering a new generation of macroscopic quantum devices, ranging from superconducting nanowire single-photon detectors operating at high temperatures\,\cite{charaev2023single, merino2023two} to twisted Josephson junctions with a tunable coupling\,\cite{zhao2023time,lee2023encapsulating,martini2023twisted}. Formation of intrinsic Josephson junctions between twisted interfaces of BSCCO thin crystals has enabled the observation of  spontaneously broken time-reversal symmetry at angles near \SI{45}{\degree}\,\cite{zhao2023time}, holding potential for realizing high-coherent superconducting quantum bits\,\cite{brosco2024superconducting}, chiral Majorana modes\,\cite{margalit2022chiral}, and novel twisted symmetries\,\cite{tummuru2022twisted}. 

\begin{figure*}[t!]
    \center
    \includegraphics[width=1\textwidth]{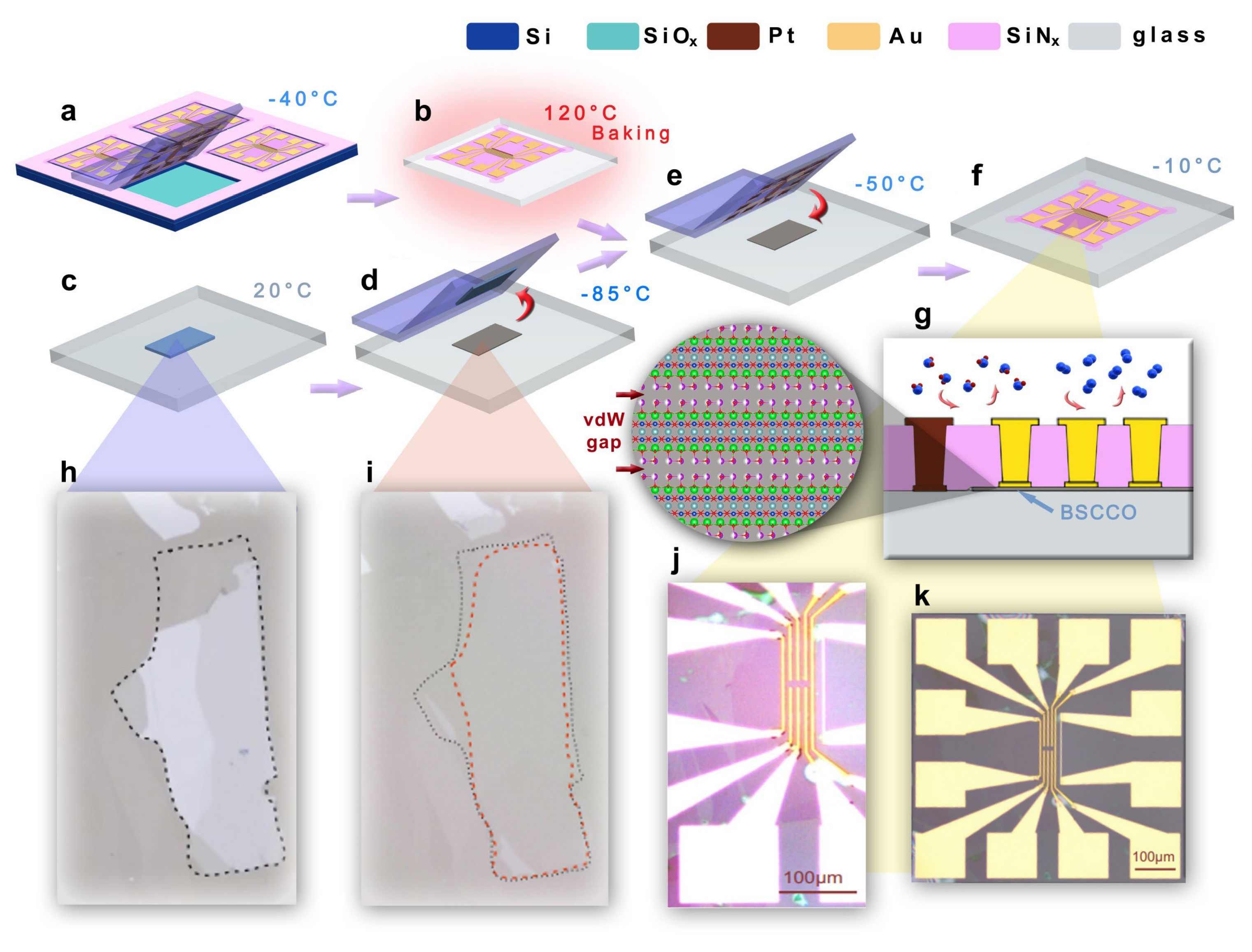}
    \caption{\textbf{Realization of atomically thin BSCCO thermoelectric circuit.} \textbf{(a-f)} Overview of the fabrication process in the Ar-filled glovebox: \textbf{(a)} Picking up the free standing NMB by PDMS at -40°C. \textbf{(b)} Baking of the NMB at 120°C to eliminate moisture. \textbf{(c)} Room temperature mechanical exfoliation of BSCCO flake onto a glass substrate. \textbf{(d)} Cryogenic PDMS technique to thin down the flake at -85°C. \textbf{(e)} Positioning of the NMB onto the glass substrate at -50°C. \textbf{(f)} Controlled release of the NMB at -10°C. \textbf{(g)} Schematic view of the NMB cross-section ion at the BSCCO flake, illustrating both the Au underlying contact adjacent with the flake and the Pt heater that is directly in contact with the glass substrate. In the zoom the crystal structure of 4.5nm, which corresponds to 1.5 u.c. of BSCCO is presented. 
    \textbf{(h-k)} Optical images of the step by step fabrication: (h) Initial exfoliated flake onto the glass substrate, underlined by a black dashed line. (i) The flake was thinned down to 1.5 u.c. by cryogenic PDMS technique, preserving the huge lateral dimensions. The red dotted line is determining the new edge of the flake. (j) Optical image in modified colors to show the position of the atomically thin flake below the NMB. (k) Overall picture of on-chip thermoelectric device.}
    \label{fig:fabrication}
\end{figure*}

Despite these promising developments, experimental investigations of vdW heterostructures are still slow  because the current technology for fabricating the vdW circuits remains limited. This occurs mostly due to the challenges of fabricating ultra-thin BSCCO crystals and integrating them into complex electrical circuits. These systems are highly susceptible to degradation from reactions with moisture and oxygen dopant loss \cite{sandilands2010stability, huang2022unveiling}. Additional challenges arise due to more subtle issues such as detrimental changes in the ordering of oxygen and this spatial ordering changes  interstitials \cite{fratini2010scale, poccia2011evolution} above \SI{-73}{\degree}C\ and spatial ordering changes of the incommensurate superlattice modulation\,\cite{poccia2020spatially}. To mitigate these issues, noninvasive optical pump-probe approaches have been opted to study the superconductivity in atomically thin BSCCO crystals\,\cite{xiao2024optically, figueruelo2024apparent}. Furthermore, in order to realize electrical measurements, stencil mask and low-power thermal evaporation have been employed for the fabrication of few electrical contacts in atomically thin BSCCO nanodevices, without involving chemicals or heating process\,\cite{zhao2019sign}. The introduction of the cryogenic exfoliation methodology allowing for picking-up and assembly of sensitive van der Waals heterostructures\,\cite{lee2023encapsulating,martini2023twisted, zhao2023time}, evidenced the potential of this technique for transformative applications\,\cite{patil2024pick}. However, the integration of the advanced circuits required to move beyond the stencil mask approach. A fabrication technique relying on the cryogenic dry transfer of printable circuits embedded into silicon nitride membranes has therefore been developed\,\cite{saggau20232d, martini2023hall}.

Ideal applications for this technology are thermoelectric circuits, requiring spatial precision at the micro-/nano-scale and a variety of elements like heaters, thermometers and electrical leads, which are often not compatible with the chemistry of the ultra-thin BSCCO crystals. Besides their natural application in energy harvesting and power generation\,\cite{zhang2022micro}, thermoelectric devices have been proven to be remarkably useful for exploring several condensed matter systems including the inert two-dimensional (2D) materials\, \cite{zuev2009thermoelectric}, topological non-trivial compounds\,\cite{gooth2017experimental}, and unconventional superconductors. In particular, the transverse thermoelectric effect and Nernst effect have been widely employed in the field of unconventional superconductivity to investigate fluctuations, symmetry breaking \cite{jotzu2023superconducting, pourret2006observation, grinenko2021state}, pseudogap phase\,\cite{daou2010broken} and vortex dynamics\,\cite{wu2024nernst, wang2006nernst, rischau2021universal}. Among the high-$T_{\text c}$ superconductors, bulk BSCCO is quite special because being extremely anisotropic\,\cite{watanabe1997anisotropic, martin1988temperature} and having large Ginzburg number and significant quantum resistance, it exhibits unique vortex transitions\,\cite{chen2007two, mounce2011charge}, which are influenced by thermal and quantum fluctuations, as well as by quenched disorder\,\cite{blatter1994vortices}. 

Learning how to control the nature of the dissipative regime can improve the performance of atomically thin cuprate superconducting circuits and the Nernst effect is a powerful tool to reach this goal. To that end, we realize silicon nitride microprinted circuit boards to perform thermoelectric measurements on atomically thin BSCCO samples and probe the vortex dynamics and dissipation regimes in presence of the effects of the sample finite sizes.
\begin{figure}[t!]
    \center
    \includegraphics[width=0.46\textwidth]{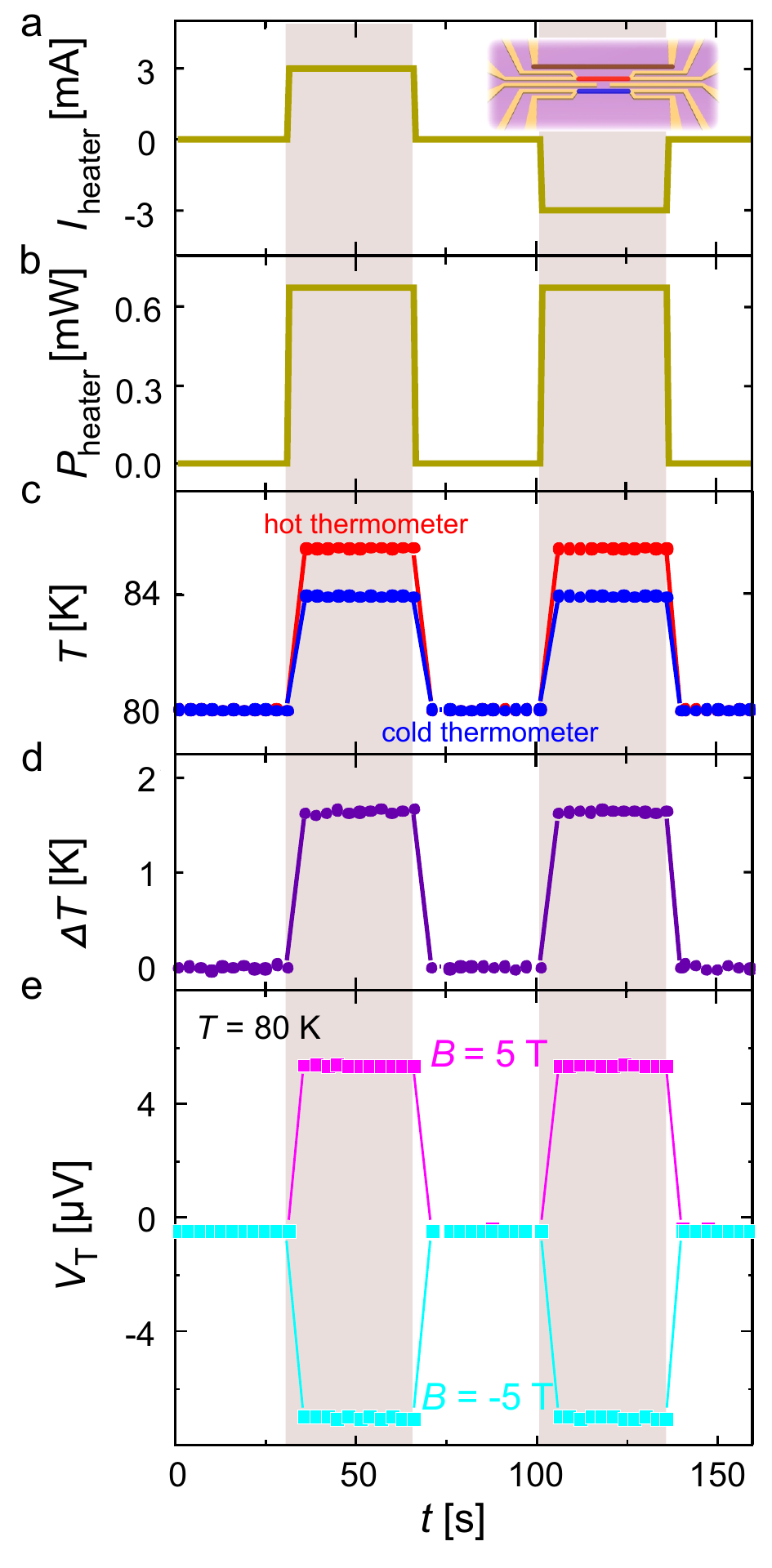}
    \caption{\textbf{Transport properties of the microprinted thermoelectric circuit.} \textbf{(a)} Current flowing in the heater as a function of time. Inset, displaying the schematics of the metallic pattern underneath the NMB shows the heater, the transverse electric contacts on the flake, and the hot (red) and cold (blue) thermometers on the substrate. \textbf{(b)} The time dependence of the power dissipated into the heater. \textbf{(c)}\,The local temperature obtained by hot and the cold thermometers as a function of time. \textbf{(d)} Time dependence of the temperature difference between the two thermometers. \textbf{(e)} Transverse thermoelectric voltage at $T$=80\,K, showing a sign change with an applied out-of-plane magnetic field $B$=$\pm$5\,T.}
    \label{fig:calibration}
\end{figure}

\section*{RESULTS AND DISCUSSION}

To realize the on-chip Nernst experimental setup, we pick up the fabricated SiN microcircuit board at -40°C using a polydimethylsiloxane (PDMS) stamp and bake it at 120°C to remove water molecules, as schematically shown in Fig.\,\hyperref[fig:fabrication]{1a,b}. The details of the circuit board fabrication are given in the Supplementary Information (SI). Following
this procedure, the BSCCO crystal with optimal doping is mechanically exfoliated onto a glass substrate within an Ar-filled glovebox (Fig.\,\hyperref[fig:fabrication]{1c}) and is then further thinned down at cryogenic temperatures using another PDMS stamp, see Fig.\,\hyperref[fig:fabrication]{1d}. This step reduces the flake thickness to the atomic level while preserving its lateral dimensions. We successfully decrease the thickness of the flake depicted in Fig.\,\hyperref[fig:fabrication]{1h} down to 4.5nm, which corresponds to 1.5 u.c., as shown in Fig.\,\hyperref[fig:fabrication]{1i}, without causing any breaks. This is possible because the glass substrate provides stronger vdW forces with the flake as compared to red forces developing in the silicon substrate used in previous studies\,\cite{lee2023encapsulating, martini2023twisted, saggau20232d}. Immediately after  that, the nanomembrane (NMB) featuring the metallic contacts underneath are transferred and freely released from the stamp on top of the flake at low temperatures, as depicted in Fig.\,\hyperref[fig:fabrication]{1e,f,g}. The quick treatment guarantees a pristine interface between the metal contacts and thin BSCCO crystal, thereby preserving its superconducting properties and ensuring an optimal electrical signal \cite{saggau20232d}. The optical micrograph of the device is shown in Fig.\,\hyperref[fig:fabrication]{1j,k}. The microcircuit consists of a resistive heater in contact with the substrate, which provides a dissipate power in the close vicinity of one side of the crystal and a set of electrodes measuring the transverse, $V$$_{\rm{T}}$, and longitudinal, $V$$_{\rm{L}}$, electric voltages.

\begin{figure*}[t!]
    \center
    \includegraphics[width=0.98\textwidth]{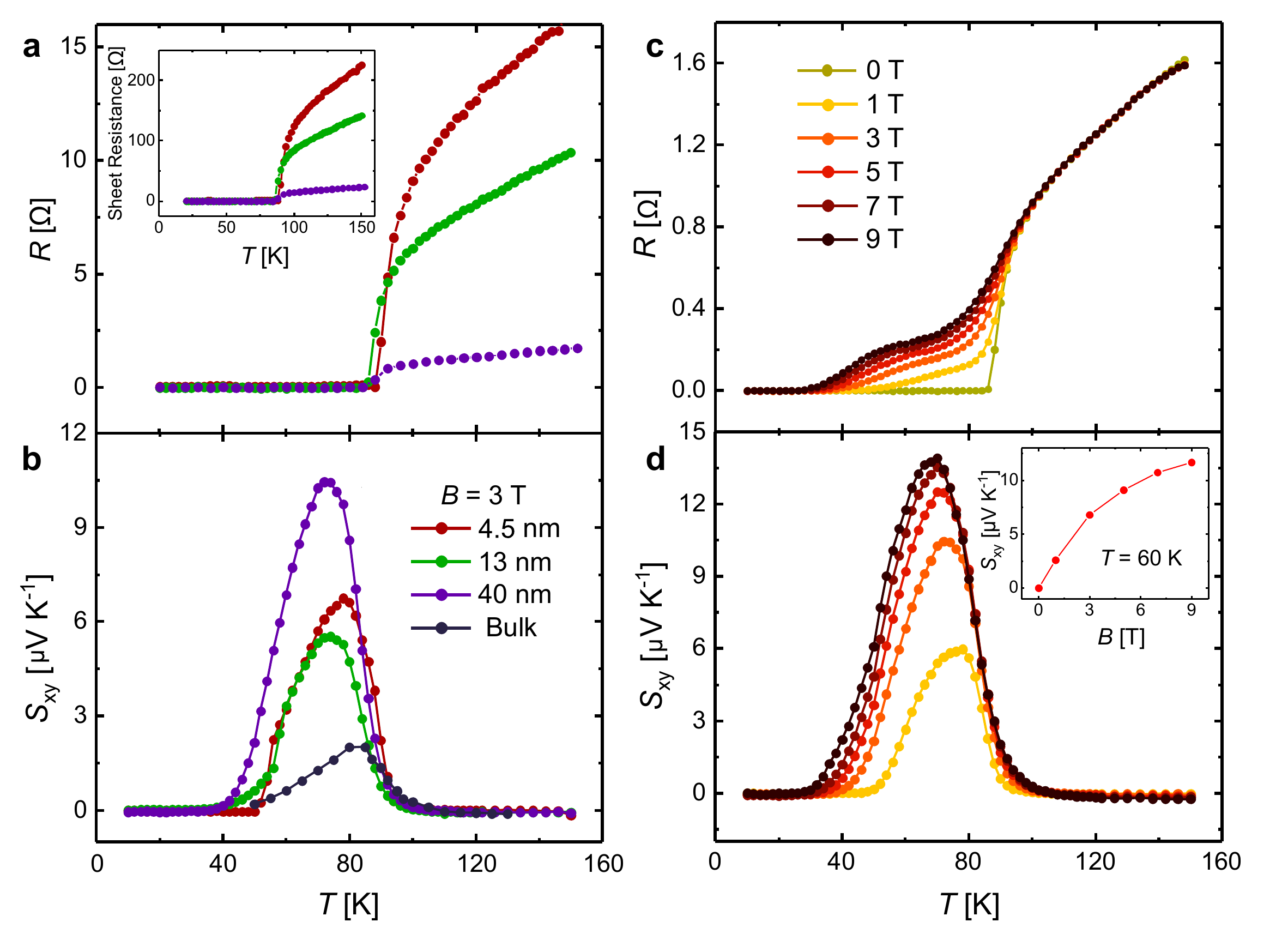}
    \caption{\textbf{Size dependent evolution of the Nernst effect.} \textbf{(a)} Four-probe resistance of the BSCCO flakes as a function of the temperature. Inset: temperature dependence of the sheet resistance for all the previous devices calculated as $\rho/t$, where $t$ is the thickness of the crystals. All samples are measured under identical geometric conditions.
    \textbf{(b)} Temperature dependence of the Nernst effect at $B$ = 3 T for BSCCO flakes in \textbf{(a)} with different thicknesses compared with optimally doped bulk adapted from\,\cite{wang2006nernst}. \textbf{(c-d)} Temperature dependence of \textbf{(c)} the resistance and of \textbf{(d)} the Nernst effect of 40 nm-thick flake under various applied magnetic fields up to 9T. Inset: magnetic field dependence of the Nernst effect at $T$ = 60 K.}
    \label{fig:measurement}
\end{figure*}

Fig.\,\hyperref[fig:fabrication]{2}, shows our measured results of the thermoelectric effect. We present the time dependence of $V_{\rm{T}}$ at $\SI{80}{\kelvin}$, when switching the heater on and off twice, as depicted in Fig.\,\hyperref[fig:calibration]{2a,b}. The two Cr/Au stripes, $20 \mu$m apart and $15 \mu$m from the heater, act as hot and cold thermometers (inset of Fig.\,\hyperref[fig:calibration]{2a}). The local temperature was inferred by four-probe resistance measurements, after a calibration of its temperature dependence from room to cryogenic temperatures (see the Fig. S3 of the SI for further details). Figure\,\hyperref[fig:calibration]{2c} presents the local temperature variation at base temperature of \SI{80}{\kelvin} as a function of time. When the heater is turned on, the local temperature increases by a few degrees, establishing a temperature difference $\Delta T$ between the two thermometers, as shown in Fig.\,\hyperref[fig:calibration]{2d}. The observed $\Delta T$ is consistent with COMSOL simulations reported in the SI. The $V_{\rm{T}}$ signal in Fig.\,\hyperref[fig:fabrication]{2e} exhibits a sign change when the external out-of-plane magnetic field of $\SI{5}{\tesla}$ is reversed, which is consistent with the expected behavior for a transverse signal that is typically an odd function of the magnetic field. This property allows us to antisymmetrize the positive and negative magnetic field data to estimate the Nernst voltage variation $\Delta V_{\rm{T}}$. Similarly, the time dependence of the Seebeck signal is reported in the SI (Fig S2). To calculate the absolute value of the Nernst coefficient $S_{\rm{xy}}$ for our thermal transport, the thermal gradient generated by the heating current must be quantified once for all the temperatures under study, see SI for further details. 

To execute our thermal transport study, we fabricate, in total, three devices for three different BSCCO crystal thicknesses: $4.5, 13, \SI{40}{\nano\meter}$. Figure\,\hyperref[fig:measurement]{3a} presents the temperature dependence of the four-probe resistance of these devices across the superconducting transition and, in the inset, the sheet resistance is presented. Notably, thinner crystals exhibit higher sheet resistance and sharper superconducting transitions, as reported in Ref.\,\cite{zhao2019sign}. In Fig.\,\hyperref[fig:measurement]{3b}, the temperature dependence of the Nernst effect is reported as a function of the thickness at $B=\SI{3}{\tesla}$ compared to the bulk one (adapted from \cite{wang2006nernst}).
In the normal state $S_{\rm{xy}}$ is small. By decreasing the temperature, a pronounced positive peak below the critical temperature appears, which corresponds to the motion of depinned vortices. Finally, $S_{\rm{xy}}$ vanishes at the
lowest temperatures. The typical signature of mobile superconducting vortices in the mixed state of cuprates \cite{wang2006nernst, behnia2016nernst} is clearly seen in all studied samples. 

\begin{figure}[t!]
    \center
    \includegraphics[width=0.45\textwidth]{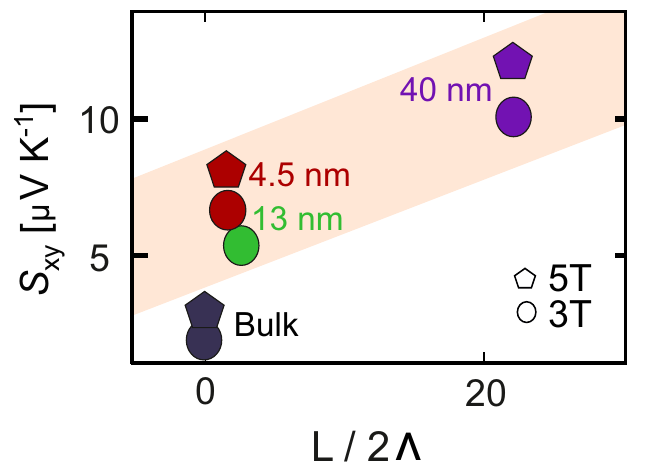}
    \caption{\textbf{Vortex distribution depending on the sample dimensions.} Experimental evolution of the Nernst coefficient $S_{\rm{xy}}$ as a function of $L$ and normalized over 2$\Lambda$
    for different magnetic fields $B$ = 5 T and $B$ = 3 T. $S_{\rm{xy}}$ is maximal in correspondence of the collective pinning regime, while it diminishes due to the effect of the finite lateral sizes. The solid black symbols related to the bulk samples \,\cite{wang2006nernst} are set as a reference for the limit of very large thickness, specifically their horizontal position is calculated as $\frac{L}{2\lambda}$, replacing $\Lambda$ by $\lambda$. }
    \label{fig:vortices}
\end{figure}

The intensity of the Nernst effect is significantly amplified in flakes as compared to that in the standard bulk samples\,\cite{wang2006nernst}, see also other experimental data in bulk BSCCO\,\cite{ri1994nernst, li2017quantitative}. Remarkably, the peak value for the $\SI{40}{\nano\meter}$-thick flake is nearly an order of magnitude greater than the peak in the bulk and twice as much as that of the $\SI{4.5}{\nano\meter}$- and $\SI{13}{\nano\meter}$-thick devices.

Interestingly, Hall effect measurements on atomically thin BSCCO flakes exhibit a sign reversal roughly in the same temperature range and the region of the $H$-$T$ plane where the Hall voltage is negative and increases its magnitude when flake thickness decreases\,\cite{zhao2019sign}. As normal excitations inside and outside vortex cores contribute to the Hall effect both below $T_{\text c}$ and above $T_{\text c}$ in the vortex-like fluctuation regime, the sign reversal indicates flux-flow dominated transport, occurring when the negative contribution from vortex cores dominates over the positive normal state contribution. The thickness dependence is attributed to the smaller mobility of the outermost atomic layers, whose relative contribution increases with decreasing thickness \cite{zhao2019sign}. However, the similarity between Hall and Nernst transverse effects cannot be extended further, as Hall effect reflects the sign of charge carriers, whereas vortex Nernst effect does not. 

To achieve a more comprehensive understanding of the work of the $\SI{40}{\nano\meter}$-thick device which exhibits the most pronounced Nernst effect, we show in Fig.\,\hyperref[fig:measurement]{3c,d} the temperature dependence of both resistance and Nernst effect as a function of magnetic field up to $\SI{9}{\tesla}$. Applying the external field broadens the superconducting transition\,\cite{ri1994nernst}, and the magnitude of the Nernst effect in the mixed state increases with field, see inset in Fig.\,\hyperref[fig:fabrication]{3d}.

The Nernst coefficient $S_{\rm{xy}}$=$E_{\rm y}/(\nabla_{\rm x}T)$ in the mixed state, where $E_{\rm y}$ is the y-component  of the electric field that results from the magnetic field's z-component $B_{\rm z}$, which is determined by the vortex dissipative motion. Thus, an enhancement of the Nernst signal indicates a growing ratio of driving to pinning forces, reflecting their competition. 
 
 Note the challenging trends of changing $S_{\rm{xy}}$ across the three flakes in our experiment: all the flakes present an enlarged Nernst coefficient with respect to the bulk sample, suggesting a crucial effect of the spatial confinement. However, among the flakes, the $\SI{40}{\nano\meter}$-one, presents the largest $S$$_{\rm{xy}}$, proving that the Nernst coeffcient is not simply scaled by the sample thickness. This evidences the fact that the vortex motion must be also influenced by the lateral confinement, giving rise to a complex scenario in which different pinning mechanisms are competing over different length scales.
It is interesting to note that our samples are all much thinner than the typical London penetration depth $\lambda$ of BSCCO, which is about 270 nm \cite{prozorov2000measurements, shibauchi1992microwave}. In this regime, the vortex-vortex interaction is mainly mediated by the magnetic stray field and the scale of the screening of the vortex in-plane supercurrents is set by the Pearl length $\Lambda$=2$\lambda^2$/$d$, rather than by $\lambda$\cite{pearl1964current}. In our samples $\Lambda$ ranges from few to tens of microns, being comparable with the characteristic lateral size  of the flakes $L$. Note that $L$ represents the smallest length of the flake. (see Table\,\ref{tab:table}). 
Thus, we attempt to represent out experimental Nernst data as a function of the ratio $L/2\Lambda$, which contains different length scales ($L$, $d$ and $\lambda$), as
shown in Fig.\,\hyperref[fig:vortices]{4}. However, though a trend is found, the role of $\Lambda$ is expected to be relevant only at very small fields, when a limited number of vortices are within the interaction range. Therefore other mechanisms, relevant at higher fields must be considered. 
In this regard, it is established that the dissipative motion of vortices across the finite-width superconducting strip involves an interplay between defects near the strip's edges, which stimulate vortex entry and exit \cite{KoshVin2002} \cite{budinska2022rising}, and the standard pinning effects of defects nearer to the strip's center, which are expected to determine vortex motion within the "depth" of the strip. In this interplay, not only the sample width, but also edge morphology and defect distribution near the edges, play key roles.

Recent reports show a systematic decrease of the Nernst effect with doping in atomically thin samples \cite{hu2024vortex}. The relevance of such effects has been demonstrated for sub-micrometric strips\,\cite{GV}. In our work, we present experimental evidence that the Nernst coefficient $S_{\rm{xy}}$ increases with the lateral width $L$ of the atomically thin samples. This observation probably indicates the growing effect of pinning in the central region of the strip (pinning effects decrease near the edges of the strip).  The control of the sample-edges geometry becomes therefore an important control knob for reducing the dissipation in  possible superconducting circuits\,\cite{brosco2024superconducting} and improving photon detectors\,\cite{charaev2023single} based on cuprate van der Waals artificial systems. This stresses the need for careful investigations of the details of pinning phenomena, particularly the possible interplay between different pinning regimes and their influence on the Nernst effect in strips of finite lateral width. 
This study, including the detailed investigation of the effects of collective pinning in the finite-width system and the possible emergence of the long range interactions, decaying as the inverse distance 1/$r$ between vortices\,\cite{pearl1964current} along with a detailed analysis of the behavior of the Nernst coefficient within the framework of pinning theory\cite{blatter1994vortices,GV,brandt2009vortex}, will be the subject of forthcoming research.

\begin{table}[t]
    \caption{Characteristic lengths of the BSCCO flakes under study }

   \begin{ruledtabular}
    \begin{tabular}{ccccc}
         $d$ [nm] & $L$ [$\mu$m] & $\Lambda$ [$\mu$m] & $L$/2$\Lambda$ & $S_{\rm{xy}}$@3T [$\mu$V K$^{-1}$ ]\\

         \hline
         4.5 &100 &32.4 &1.54 &6.74  \\ 
         13 &56 &11.2  &2.50 &5.53 \\ 
         40 &155 &3.65 &22.3 &10.44  
    \end{tabular}
    \end{ruledtabular}
    \label{tab:table}
\end{table}

\section*{Experimental section}
\noindent
\textbf{Device fabrication}: The exfoliated flakes were obtained within an Ar-filled glovebox from a bulk single crystal of BSCCO and transferred from the scotch tape on a glass substrate, i.e. quartz monocrystals with various different crystal directions formed via hydrothermal synthesis. The substrate has been both pre-treated with oxygen plasma to optimize the flake adhesion and placed on a hotplate at 150°C for several hours to remove the moisture. The thicknesses of the flakes were estimated, based on the RGB calibration by the optical microscope. Nevertheless, the actual thickness was accurately determined using atomic force microscopy (AFM) after the completion of the transport measurements. To successfully perform the cryogenic thinning described in the main text, the glass substrate was placed on the stage and fixed at --85°C. This temperature, which is low enough to freeze the interlayer O$_2$ dopants, has been chosen based on the dew point of the glovebox, --97°C. Note that the cryogenic PDMS treatment at --85°C was employed to remove thick flakes adjacent to the target flake, for arranging the optimal NMB landing. The precise alignment of the NMB at  
--50°C\,with the flake was facilitated by a motorized micro manipulator.
After removing the PDMS at --10°C, The device reached room temperature before removing from the glovebox. Finally, the glass substrate was affixed to the chip carrier using silver paste and wire bonded with the automatic BONDTEC 5630i.
\medskip
\noindent
\textbf{Transport measurements}: The transport measurements were performed in a 9 T Quantum Design Physical Property Measurement System (PPMS). For the resistivity measurements, a standard four-probe method was applied using external SR830 lock-in amplifiers.
For the on-chip Nernst effect measurements, we evaluated both the transverse thermoelectric voltage, $\Delta V_{\rm T}$, and the temperature difference across the flake, $\Delta T$.
First, $\Delta V_{\rm{T}}$ was obtained for all the temperatures under study, following the procedure described in the main text, see Fig.\,1, and performing an antisymmetrization of two curves at positive and negative magnetic fields. Then, the calibration of the temperature gradient from room to low temperature was done once with the particular patterned NMB reported in Fig.\,2. A calibration of the four probe resistance of the hot and cold thermometers was done to evaluate the local temperature by four-probe resistance measurements, see Fig.\,2f,g. All thermoelectric measurements were performed using an external current source KEYTHLEY 4200 and nanovoltmeters KEYTHLEY 2182A. All the samples were measured with identical NMBs. 
\vspace{.2cm}

\noindent \textbf{Acknowledgments--}
The work is partially supported by the Deutsche Forschungsgemeinschaft (DFG 512734967, DFG 492704387, and DFG 460444718), co-funded by the European Union (ERC-CoG, 3DCuT, 101124606). The work of V.M.V. was supported by Terra Quantum AG. The work at BNL was supported by the US Department of Energy, office of Basic Energy Sciences, contract no. DOE-sc0012704. The authors are  grateful for technical support to Ronny Engelhart, Christiane Kranz, Ronald Uhlemnann and Nicolas Rodriguez Perez. The authors thank Heiko Reith and Dietmar Berger for the technical support on the COMSOL simulations. The authors are deeply grateful to  Francesco Tafuri, Haider Golam, Uri Vool, Luca Chirolli, Valentina Brosco, Domenico Montemurro, and Davide Massarotti for support and illuminating discussions.

\vspace{0.2 cm}
\noindent \textbf{Author contributions--}
N.P. conceived and designed the experiment; S.S, performed the fabrication. C.N.S. and S.S. initiated the nanomembrane fabrication. S.S. and M.C. performed the measurements and analyzed the data with the contribution of F.C. and I.P. The cuprate crystals were provided by G.G. The fabrication procedure were discussed by S.S., M.C., T.C., M.M., Y.L., K.N., N. P., and the results were discussed by F.C., I.P., K.N., N.P., V.M.V. The manuscript was written by S.S., M.C., M.M., V.M.V., I.P., K.N., F.C., N.P.  All authors discussed the manuscript.

\vspace{0.2 cm}
\noindent \textbf{Competing interests--}
The authors declare no financial or non-financial competing interests.


\bibliographystyle{}

\end{document}


\newpage
\textbf{1. Nanomembrane fabrication}

\begin{figure*}[h!]
    \center
    \includegraphics[width=1\textwidth]{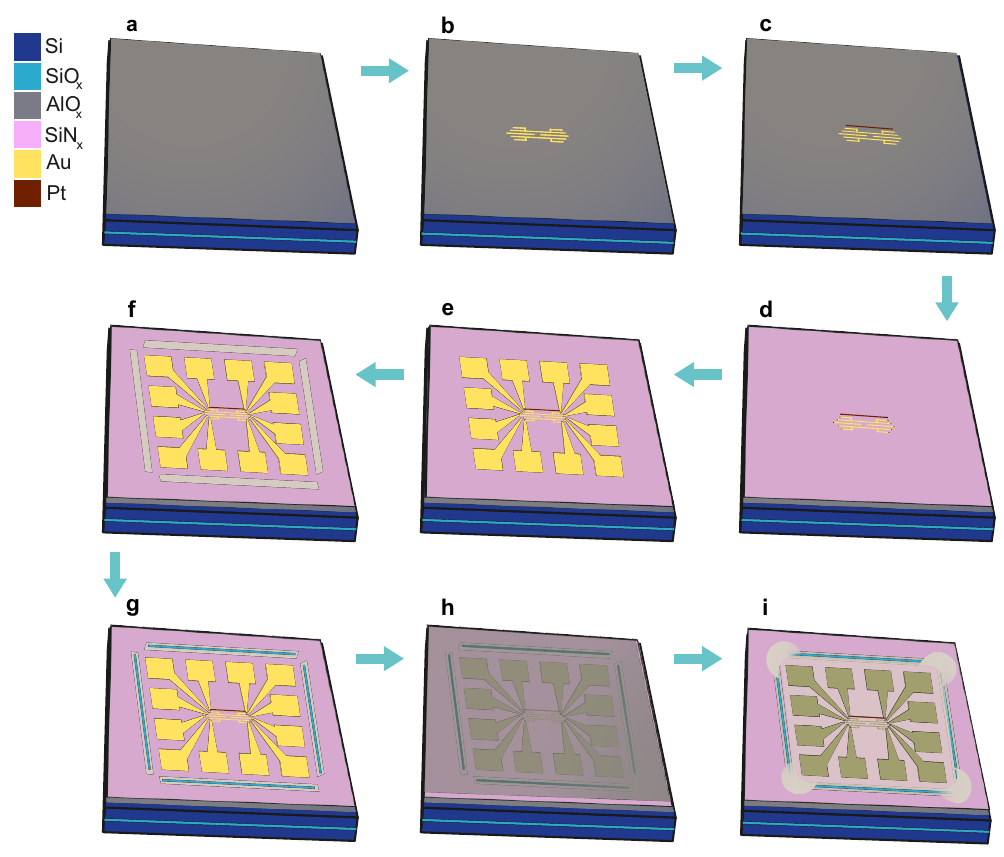}
    \caption{\textbf{Thermoelectric NMB fabrication process.} a) Deposition of a 2-3 nm sacrificial Al$_2$O$_3$ layer by ALD on the SiO$_2$/Si/SiO$_2$ substrate. b) Au deposition to establish underlying contacts for the thermoelectric configuration. c) Deposition of Pt/Ti stripes as a heater. d) CVD deposition of 800 nm SiN$_x$ followed by via etching. e) Deposition of Cr/Au to create large pads for subsequent wire bonding, connecting to underlying contacts. f) Etching through SiN$_x$ g) followed by etching through the first Si layer inside the big stripes to separate the NMB from each other h) Second Al$_2$O$_3$ deposition to passivate the membrane for subsequent step. i) XeF$_2$ etching to remove the Si layer and produce the free-standing NMB. }
    \label{fig:NMB fabrication}
\end{figure*}

The fabrication of the nanomembrane (NMB) to measure the thermoelectric properties started from the usage of a pristine Si/SiO$_2$/Si (10/1/2.5 $\mu m$) (silicon on insulator, SIO) substrate subjected to a 40-minute O$_2$ plasma treatment to ensure cleanliness. Subsequently, to ensure the final achievement, a 3 nm Al$_2$O$_3$ sacrificial layer was deposited through atomic layer deposition (ALD) from Oxford Company (Fig. \ref{fig:NMB fabrication}a). Metal contacts were established (Fig. \ref{fig:NMB fabrication}b,c) by photo-lithography in two steps to sputter Au (80 nm) for the electric contacts and Ti/Pt (4/80 nm) stripes serving as a heater separately. The deposition of the SiN$_x$ layer was accomplished using plasma-enhanced chemical vapor deposition (PECVD) (Fig. \ref{fig:NMB fabrication}d). Both the thickness and the elasticity of the SiN$_x$  layer could be controlled. Before creating the large bonding pads intended for wire bonding by lithographic process and sputtering of Cr/Au (4/80 nm), an application of reactive ion etching (RIE) was leveraged, resulting in the precise etching of holes atop the lower gold contact to create a seamless connectivity between the upper and lower gold contacts (Fig. \ref{fig:NMB fabrication}e). Each thermoelectric NMB with 12 pins had dimensions of 750 $\mu$m $\times$ 750 $\mu$m, and 49 NMBs could be fabricated per cm$^2$ of SIO. Therefore, each NMB needed to be isolated from its neighboring NMBs. To achieve this, deep reactive ion etching was performed at -120$^\circ$C to etch through the Si layer and reach the SiO$_2$ layer (Fig. \ref{fig:NMB fabrication}f,g). To release the SiN$_x$ nanomembrane by elimination of the underlying Si layer, XeF$_2$ gas was employed. Before this step, to passivate the NMB from the fluorine etching, another layer of 3 nm of Al$_2$O$_3$ was deposited on the clean samples (Fig. \ref{fig:NMB fabrication}h). Subsequent to the Si etching, the Al$_2$O$_3$ sacrificial layer was etched using MIF to make free standing NMB. Then, it was immersed in isopropanol overnight before critical point drying (CPD). The step-by-step details of the NMB fabrication process are described in the work of Saggau et al. \cite{saggau20232d}.

\clearpage
\newpage
\textbf{2. Seebeck effect of atomically thin BSCCO flake}

\begin{figure}[h!]
    \center
    \includegraphics[width=0.7\textwidth]{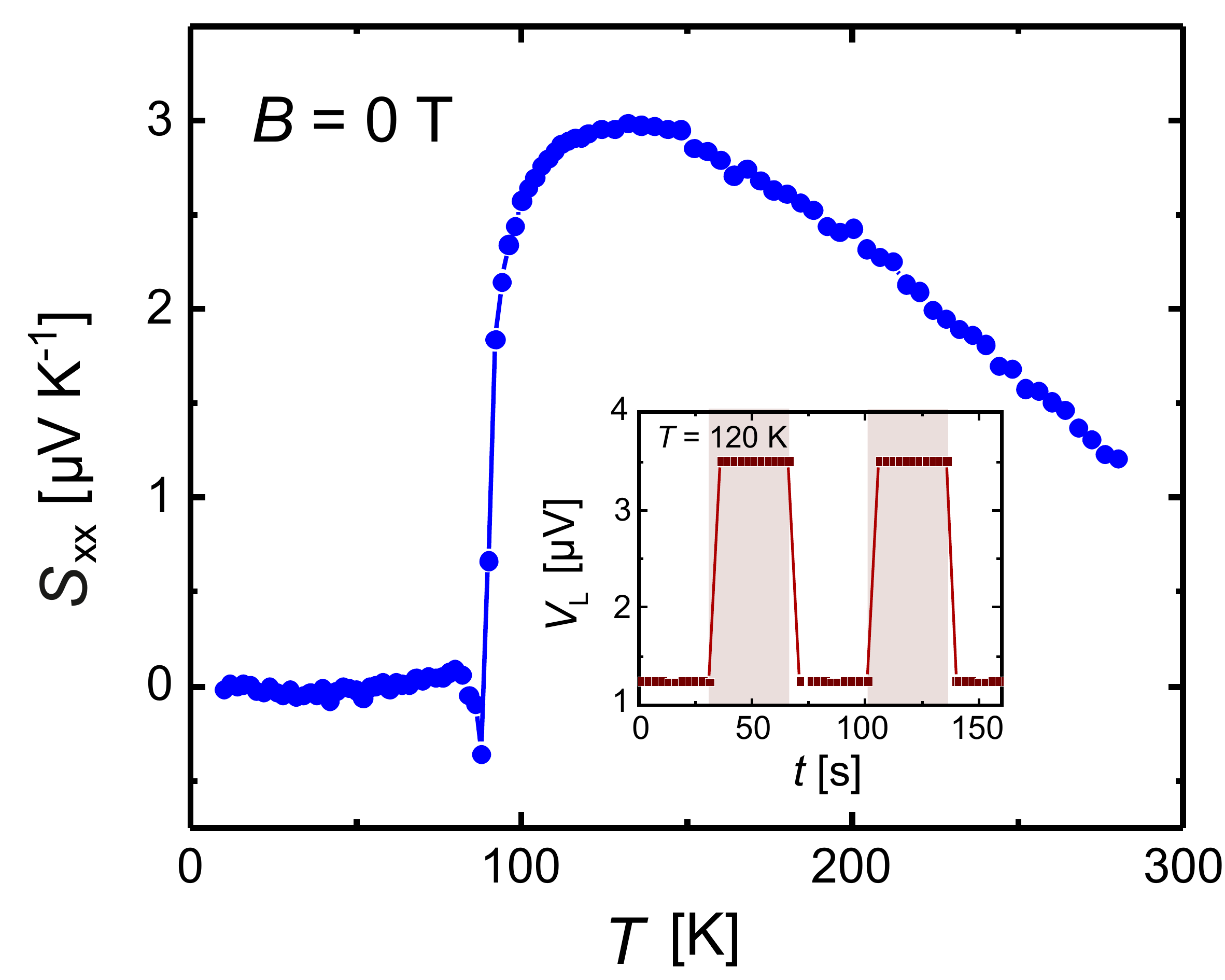}
    \caption{\textbf{Seebeck effect as a function of temperature of 1.5 u.c.} BSCCO flake in zero magnetic field conditions. Inset, longitudinal thermoelectric voltage at $T$=120 K in the absence of magnetic field. The heater was switched on twice in the shaded regions.}
    \label{fig:Seebeck}
\end{figure}

Figure \ref{fig:Seebeck} illustrates the temperature dependence of the Seebeck effect for the 1.5 u.c. flake, across the superconducting transition. Both the Seebeck voltage and the temperature gradient were measured following the procedure detailed in the main text (Inset of Figure S2).

The Seebeck effect ($S_{\rm{xx}}$) exhibits the typical behavior of bulk BSCCO across the superconducting transition \cite{ri1994nernst}. Above $T_{\rm{C}}$, $S_{\rm{xx}}$ increases as the temperature decreases, followed by a sudden drop at the superconducting transition. Below $T_{\rm{C}}$, the Seebeck effect becomes zero. The inset in Figure S2 is the time dependence
of the $V_L$ at $T$ = 120 $K$, switching on
and off the heater two times.

\clearpage

\textbf{3. Calibration of the thermal gradient}

To calibrate the thermal gradient created by the on-chip thermoelectric device, a specific NMB, equipped with two thermometers and a heater, was transferred onto a pristine glass substrate (Quartz monocrystals with various different crystal directions formed via hydrothermal synthesis). The two metallic stripes, acting as thermometers, were $20 \mu$m apart and $15 \mu$m from the heater. The process mirrored the device fabrication described in the main text: the NMB was picked up at -40 °C using a cryogenic PDMS technique shown in  Fig. \ref{fig:Seebeck}a and placed onto the glass at the same temperature (Fig. \ref{fig:Seebeck}b). Subsequently, the PDMS was removed from the substrate at -10 °C (Fig. \ref{fig:Seebeck}c). The device was then bonded outside the glove box and loaded into the cryostat to perform the 4-probe R vs T measurement for thermometer calibration. 
\vspace{2cm}

\begin{figure}[h!]
    \center
    \includegraphics[width=1\textwidth]{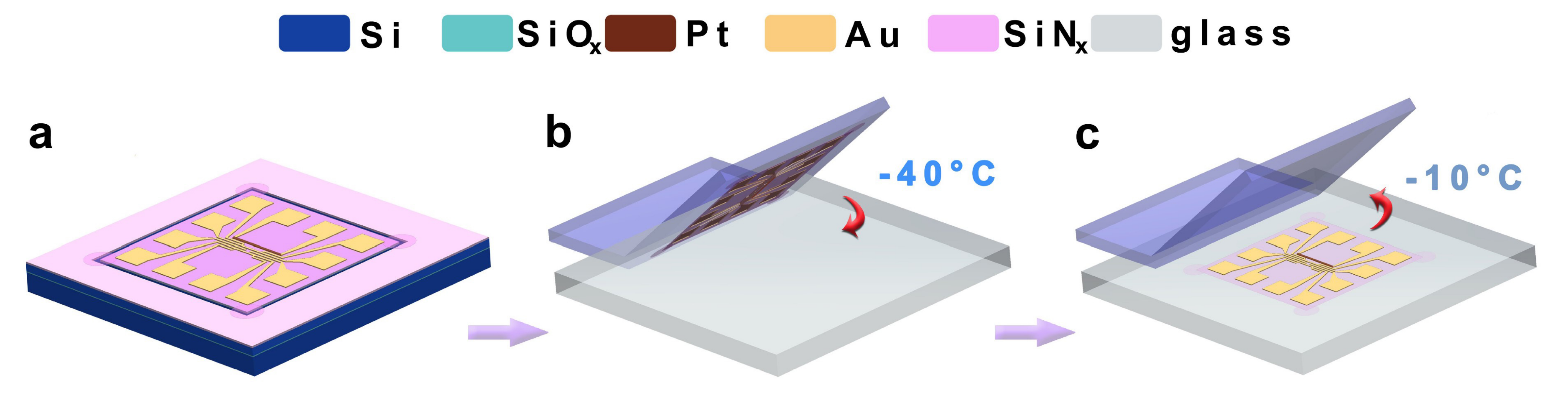}
    \caption{\textbf{Overview of the device fabrication to calibrate the thermal gradient.} (a) Free standing NMB (b) the pick up of the NMB by PDMS at -40 °C and (c) to the transfer of the NMB onto the glass substrate at -10 °C}
    \label{fig:fab_th}
\end{figure}

The thermometers were made by metallic stripes and the temperature dependence of the resistivity was showing the standard metallic behavior. Hence, for a given $T$, the calibration of the temperature dependence of the resistivity was allowing to unequivocally associated a local temperature to the measured R value. The evaluation of the temperature gradient from room to low temperature was done once, following the procedure described in the Fig. 2 of the main text. The calibration was possible only avoiding the low temperature region, approximately below 10 K, in which R was showing a saturation.

\textbf{4. COMSOL simulations of the themperature gradient}

To properly modelize the temperature gradient across the flake, a standard COMSOL simulation of the experimental set-up (Fig. 2d in the main text) was performed.
In Figure S3a the schematic of the simulated experimental set-up is displayed. A 800nm-thick SiN$_x$ NMB with the Pt heater underneath was positioned onto the glass substrate (Fig. 2d in the main text).

\begin{figure*}[h!]
    \center
    \includegraphics[width=1\textwidth]{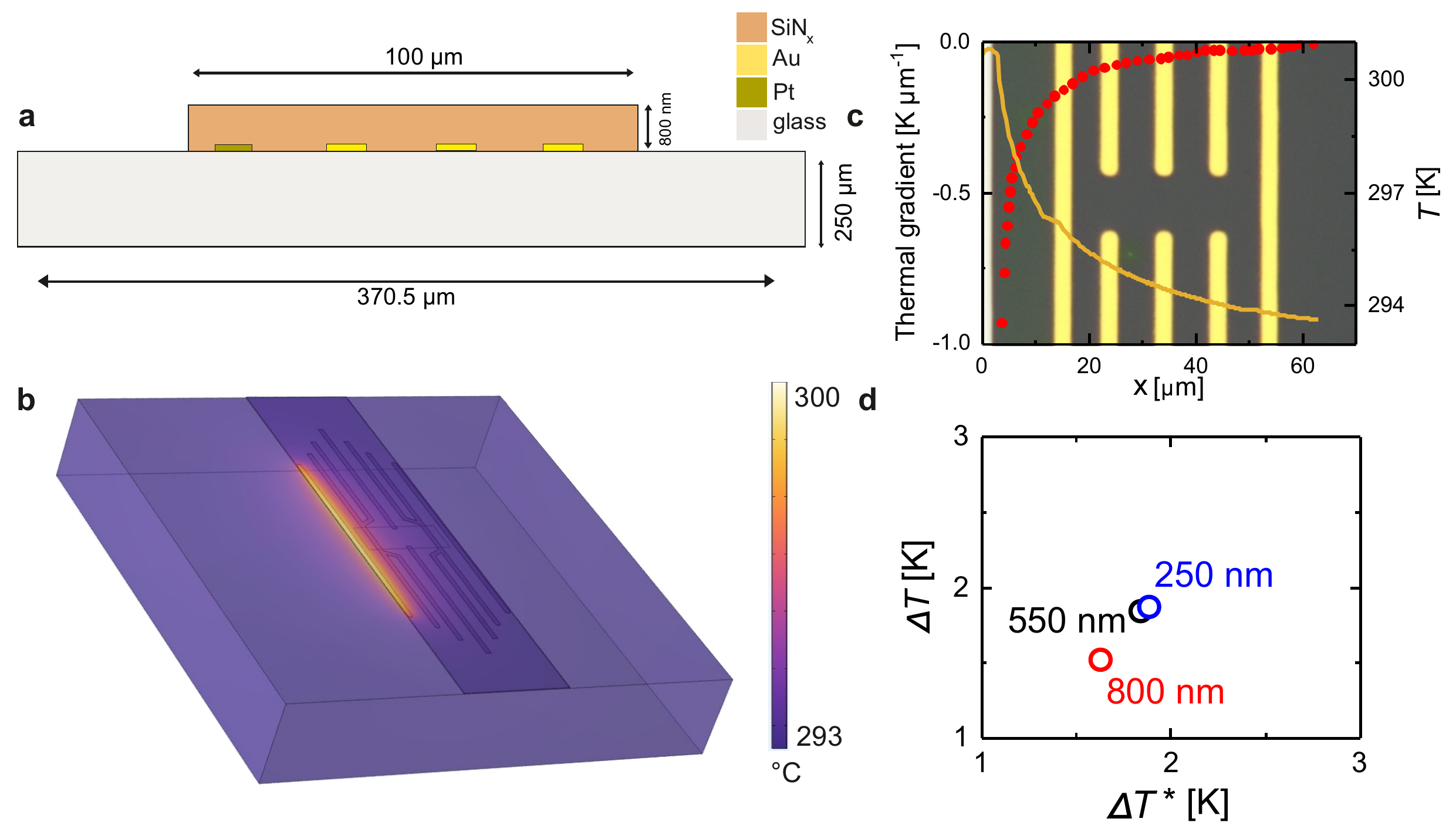}
    \caption{\textbf{COMSOL simulations of the temperature distribution} a) Schematic of the experimental set-up: 800nm-thick NMB with the Pt heater underneath positioned onto the glass substrate. b) Simulated temperature distribution close to the heater, applying 3 mA to the heater at 293.5 $K$. c) Simulated local temperature and thermal gradient, respectively in orange and red, as a function of the distance from the heater. The image of the pattern for thermoelectric transport measurements is superimposed to the graph. d) Simulated temperature difference between hot and cold thermometers as a function of the change of the mean temperature between them for 250 nm, 500 nm, 800 nm thick NMBs.}
    \label{fig:simulation}
\end{figure*}

Figure \ref{fig:simulation}b, c present the simulated local temperature and temperature gradient close to the heater, applying 3 mA in the heater at 293.5K using the parameters as shown in \ref{fig:simulation}a. As already described in the main text, the local temperature increases by a few degrees ($\Delta T^*$) when the temperature difference between the two thermoelements ($\Delta T$) is established (Fig. 3h in the main text). Although the temperature gradient is enhanced in the heater region, it is quite uniform in the flake region. At 293.5 $K$, the simulated $\Delta T \approx 1.5 K$ is reasonable compared to experimental values. Furthermore, the results of the simulations are not changing much varying the thickness of the NMB from 800 nm (experimental configuration) to 250 nm, underlining that the temperature gradient is mainly established into the substrate as it is shown in Fig. \ref{fig:simulation}d.

\clearpage
\bibliographystyle{}